\documentclass[onecolumn,showpacs,showkeys,preprintnumbers,amsmath,amssymb]{revtex4}

\usepackage{graphicx}% Include figure files
\usepackage{dcolumn}% Align table columns on decimal point
\usepackage{bm}% bold math

\begin{document}

%\preprint{}
\title{On the functional form of the infinite square well model}

\author{Chyi-Lung Lin}
\email{cllin@scu.edu.tw}
%\email[]
%\thanks{}
\affiliation{Department of Physics,\\ Soochow University, \\Taipei,Taiwan, R.O.C.}

%\author{\thanks{ E-mail: \email{ }}}

%\institute{Department of Physics,\\ Soochow University, \\Taipei,Taiwan, %R.O.C.}

%\pacs{03.65.w}{ Quantum mechanics}


%\date{\today}

\begin{abstract}
{ 

The original model of the infinite square well contains a vague notation $\infty$ and therefore results some ambiguities. We investigate to obtain a functional form for the potential energy $V(x)$. This is done by substituting back the original energy eigenstates and eigenvalues into the Schr\"odinger equation. We then obtain a precise functional form of the $V(x)$. 
From this reformed model, we show that energy eigenstates and eigenvalues can directly be obtained without the need of imposing boundary condition, Ehrenfest's theorem can directly be confirmed, and ambiguities in the original model can be resolved. 

}

\end{abstract}

%\keywords{}



%----------------------------------

\maketitle


\section{\label{sec:INT}  Introduction}

The infinite square well is a model using infinitely high potential barrier to confine particles inside a well. We also have other models using infinitely large potential energy, such as the Dirac delta function potential. Recently, Belloni and Robinett have given a review on these models \cite{b.r1}. 
The potential energy $V(x)$ of the infinite square well is described by
\begin{equation}
\label{Eq.INT-1}
V(x)=\left\{
\begin{array}{ll}
0,&        0<x<L, \\
 \infty,&      \text{otherwise}. 
\end{array}
\right.
\end{equation}
The notation $\infty$ is used in  Eq.~(\ref{Eq.INT-1}). Such a form of potential energy does not have a precise functional form.
In what follows, we investigate to reform above  potential energy in terms of established functions so that $V(x)$ has a precise functional form,
and therefore avoid some ambiguities which will be discussed below.
%We can then avoid using the notation of $\infty$, 

The time independent Schr\"odinger equation  is
\begin{equation}
\label{Eq.INT-2}
-\frac{\hbar^2}{2m} \Psi''(x)+V(x) \Psi(x)=E \Psi(x).
\end{equation} 
For $V(x)$ defined in  Eq.~(\ref{Eq.INT-1}), 
the energy eigenfunctions $\Psi_n (x)$ and the eigenvalues $E_n$ of Eq.~(\ref{Eq.INT-2}) are well-known \cite{b.r2,b.r3,b.r4,b.r5, b.r6,b.r7,b.r8}. We have
\begin{equation}
\label{Eq.INT-3}
\Psi_n (x)=\left\{
\begin{array}{ll}
\sqrt\frac{2}{L}\ \sin(k_n  x),&        0<x<L, \\
  0,&      \text{otherwise}.
\end{array}
\right.
\end{equation}
\begin{equation}
\label{Eq.INT-4}
E_n=\hbar^2  k_n^2/ 2 m,
\end{equation}
where $k_n= n \pi/L$, $n=1,2,3...$.  
%%%%
%%%%%
These solutions are obtained by imposing boundary conditions at the two sides of the well.  
The boundary condition imposed is that we require and only require the continuity of  $\Psi(x)$ at the two sides of the well. This then yields the solutions described in Eqs.~(\ref{Eq.INT-3}-\ref{Eq.INT-4}). If we also impose the continuity of $\Psi'(x)$, then we only obtain a trivial solution  $\Psi(x)=0$. 
We can only impose boundary conditions, because the form of the $V(x)$ in Eq.~(\ref{Eq.INT-1}) does not allow us to determine the boundary condition from Schr\"odinger equation. \cite {b.r7, b.r8}.  

Although we have obtained solutions by imposing boundary conditions, nevertheless, it seems dangerous in physics using a quantity such as infinity. There is no controversy over the delta function potential, because we know how to handle the infinity; we have the formula as

\begin{equation}
\label{Eq.INT-5}
\int_{-\infty}^\infty   \delta(x) dx = 1.
\end{equation}
This is a specification formula showing how to handle the infinity. And this results that boundary conditions can be derived from Schr\"odinger equation for the delta function potential. 
In contrast, the quantity $\infty$ in Eq.~(\ref{Eq.INT-1}) is quite a vague notation. It lacks a specification formula like that in Eq.~(\ref{Eq.INT-5}). We therefore have no guide on how to deal with it. And for this reason, we have to impose extra boundary conditions to obtain solutions.
%for the model defined in Eq.~(\ref{Eq.INT-1}). 
Also, there are ambiguities resulted from straightforwardly using this notation of $\infty$. 
These are described as follows:

(1) The first ambiguity is: what is the value of $ V(x) \Psi_n(x)$ outside the well, where $V(x)= \infty$ and $\Psi_n(x)=0$?  We would encounter the problem as  $ V(x) \Psi_n (x)=\infty \times 0 = ?$ As $E\Psi(x)=0$ and $\Psi''(x)=0$ outside the well, therefore from the Schr\"odinger equation, Eq.~(\ref{Eq.INT-2}), we need $ V(x) \Psi_n (x)=\infty \times 0 = 0$.

(2) The second ambiguity is: what is the value of $ V(x) \Psi_n(x)$ at the sides of the well, where again $V(x)= \infty$ and $\Psi_n(x)=0$? However, in this case, we need $ V(x) \Psi_n (x)=\infty \times 0 = \infty$. 
This had already been pointed out that  $V(x) \Psi_n(x)$  should have a delta function at each side of the well \cite {b.r9, b.r10}. 
%This is because the value of $V(x) \Psi_n(x)$ at the side is related to the discontinuity of $\Psi'_n(x)$. 
%
%

(3) The third ambiguity concerns the manifestation of Ehrenfest's theorem.
We note that the term $-dV(x)/dx$ represents a force, which is zero inside the well but is infinite at the sides. Yet, the probability density $\Psi^{*}(x) \Psi(x)$ is zero at the two sides.  Hence, calculating the expectation value of the force, we encounter the problem  at the edges that $dV(x)/dx\ \Psi^{*}(x) \Psi(x) =\infty \times 0 = ?$ \citep{b.r11}.

We have different ways for solving these ambiguities. The first is an indirect way;  that is, we consider the infinite square well potential as a limiting case of a finite well potential  \citep{b.r10, b.r11}. 
Seki and Rokhsar calculated quantities in a finite well, and followed by taking the limit  $V_0 \to \infty$. In this way, we can calculate the value of $V(x) \Psi(x)$ and also the expectation values to confirm Ehrenfest theorem in the infinite square well \citep{b.r10, b.r11}. 
%
%The energy eigenstates and eigenvalues of a finite well with depth $V_0$ are shown in the Appendix I. 
%

The second is a direct way. The above ambiguities shows that we need a more precise functional form for the potential energy. Therefore, it  is interesting and worth to investigate the limit of the functional form of the potential energy. 
This is quite similar to the definition of Dirac delta function $\delta(x)$, which usually is defined as
\begin{equation}
\label{Eq.INT-6}
\delta(x)=\left\{
\begin{array}{ll}
0,&       x \neq 0, \\
 \infty,&      x=0. 
\end{array}
\right.
\end{equation}
We may consider it as the limit of the finite square step function, defined as: 

\begin{equation}
\label{Eq.INT-7}
f(x, \epsilon)=\left\{
\begin{array}{ll}
\frac{1}{\epsilon},&   \frac{-\epsilon}{2}< x <\frac{\epsilon}{2}, 
\\
 0,&      \text{otherwise}. 
\end{array}
\right.
\end{equation}
Simply taking the limit $\epsilon \to 0$ leads directly to Eq.~(\ref{Eq.INT-6}). But, on the hand, we do have a functional form to this limit, that is 
\begin{equation}
\delta(x)= \frac{1}{2 \pi} \int_{-\infty}^\infty e^{- i x y} dy.
\label{Eq.INT-8}
\end{equation} 
This is a very useful formula for describing the delta function. 

In the same spirit, we seek a functional form of $V(x)$ for the original infinite square well model. 
%
%We can then directly discuss the infinite square well with this reformed $V(x)$. 
%
%
The potential energy of a finite well may also be described as: $V(x)=V_0 [\theta(-x)+ \theta(x-L)]$. Taking the limit $V_0 \to \infty$ simply leads to the $V(x)$ in Eq.~(\ref{Eq.INT-1}). 
This is not the functional form we are looking for.
%This is just a mathematical approach, in fact, we do not expect there is a functional form for this limit. 
Instead, we are looking for an approach to study the limit of a finite square well physically.
%We ask what is the functional form of a potential energy that results the solutions described in Eqs.~(\ref{Eq.INT-3})-(\ref{Eq.INT-4}).
%Thus, the limit of the functional form of $V(x)$ should be guided by Schr\"odinger equation. 

%
%

To explore such a  functional form physically, we work in an opposite way. That is, from $\Psi(x)$ to determine  $V(x)$.   
By taking $V(x)$ as an unknown, and substituting back a wave function $\Psi(x)$ into the Schr\"odinger equation, we can then determine the corresponding potential energy. We confirm this method in the Appendix I for the case of the finite well.

For the infinite square well, the wave function in Eq.~(\ref{Eq.INT-3}) can be rewritten in a more compact form as 
\begin{equation}
\label{Eq.INT-9}
\Psi_n(x)= \sqrt \frac{2}{L} \sin(k_n  x) \theta(x)\theta(L-x).
\end{equation}
The term $\theta(x)\theta(L-x)$ can also be expressed as $[ \theta(x)-\theta(L-x)]$   \cite{b.r1}.
We then substitute this form of $\Psi_n(x)$  into the Schr\"odinger equation, Eq.~(\ref{Eq.INT-2}). 
This then yields
\begin{equation}
\label{Eq.INT-10}
V(x) \Psi_n (x)
=   \sqrt {\frac{2}{L}} \frac{\hbar^2}{2m}  k_n \Big[  \delta(x) -  \cos(k_n L)  \delta(L-x) \Big].
\end{equation}
The derivation is shown in Sec. II. Eq.~(\ref{Eq.INT-10}) shows the required delta functions that are inherent in the values of $V(x) \Psi_n(x)$   \cite {b.r9, b.r10}.

Eq.~(\ref{Eq.INT-10}) is an important result. We show in Sec. IV that  Eq.~(\ref{Eq.INT-10}) enables us to directly resolve all the ambiguities mentioned above. That is, we need not solve these ambiguities by means of finite well.
More importantly, Eq.~(\ref{Eq.INT-10}) shows the property of the functional form of the $V(x)$ in the infinite square well.
The $V(x)$ containing the notation $\infty$ in Eq.~(\ref{Eq.INT-1}) is now replaced by a functional form.
If we are to confine particles inside a well with $\Psi_n(x)$ as the energy eigenstates, then the potential energy should have such a property described in Eq.~(\ref{Eq.INT-10}).
We  note that $\Psi_n(x)$ contains the factor $\theta(x) \theta(L-x)$. Because the divergence nature in the $V(x)$ of the infinite square well, this factor shows how to handle the infinity in the $V(x)$. From this formula, we can go on to investigate a more precise functional form of $V(x)$.

In Sec. II, we show the derivation of the functional form of $V(x)$ from the known  eigenfunctions  $\Psi_n(x)$ and eigenvalues $E_n$. 
We call the system defined by this $V(x)$ the reform model. 
In Sec. III, we show that the energy eigenstates and eigenvalues of the reformed model are the same as those in the original model. 
In Sec. IV, we show that above ambiguities can be resolved by the reformed model.
In Sec. V, we have a conclusion.
%%%
%%%
In Appendix I, we show that the functional form of the potential energy of a finite well can be obtained by the solutions of the finite well. 

%In Appendix II, we discuss more about the functional form of the potential energy $V(x)$. 

\section{\label{sec:TDE} The derivation of the functional form of $V(x)$ from $\Psi_n(x)$}

We here show the derivation of the functional form of $V(x) $  from the energy eigenstates $\Psi_n(x)$ descried in Eq.~(\ref{Eq.INT-6}).
As we are to derive the form of $V(x)$ from the wave function,  we then rewrite the Schr\"odinger equation as:
\begin{equation}
\label{Eq.TDE-1}
V(x) \Psi(x) = \frac{\hbar^2}{2m} \Psi''(x)+E \Psi(x).
\end{equation} 
The $\Psi''(x)$ can be calculated by the following formula
\begin{eqnarray}
&& \Psi(x)=D(x) \theta(x) \theta(L-x)
\nonumber\\
&&\Psi''(x)=D''(x) \theta(x) \theta(L-x) + D'(x) \delta(x)  - D'(x)  \delta(L-x) .
\label{Eq.TDE-2}
\end{eqnarray}
Substituting this formula of $\Psi''(x)$ into the right side of Eq.~(\ref{Eq.TDE-1}) 
and replacing  $\Psi(x)$ by $\Psi_n (x)$, energy $E$  by $E_n$, 
and $D(x)$ by $D_n(x)$ where
\begin{equation}
D_n(x)=\sqrt \frac{2}{L} \sin(k_n  x),
\label{Eq.TDE-3}
\end{equation}
this then yields
\begin{equation}
V(x) \Psi_n (x)
=
 \frac{\hbar^2}{2m}   \Big[  \delta(x) -   \delta(L-x) \Big] D_n'(x).
\label{Eq.TDE-4}
\end{equation}
Such a form is not symmetrical with respect to the two edges. We need a more symmetrical one. Substituting Eq.~(\ref{Eq.TDE-3}) into Eq.~(\ref{Eq.TDE-4}), we have 
\begin{eqnarray}
V(x) \Psi_n (x)
&=& \sqrt {\frac{2}{L}}  \frac {\hbar^2}{2m} \ k_n \cos(k_n x) \Big[ \delta(x) 
-    \delta(L-x) \Big],
\nonumber\\
&=&    \sqrt {\frac{2}{L}} \frac{\hbar^2}{2m}  k_n \Big[  \delta(x) -  \cos(k_n L)  \delta(L-x) \Big].
\label{Eq.TDE-5}
\end{eqnarray}
This is the formula we state in Eq.~(\ref{Eq.INT-10}). 
In what follows, we try to obtain a functional form of $V(x)$ from Eq.~(\ref{Eq.TDE-5}). 
We need to rearrange the right side of Eq.~(\ref{Eq.TDE-5}). 
Using the result that $\lim_{x\to 0} \sin (k_n x)/x =k_n$, the first term,  
$\sqrt{2/L} \  k_n \delta(x)$, can simply be written as $[D(x)/x] \delta(x)$. We also need to rewrite the second term, $-\sqrt{2/L} \  k_n \cos(k_n L) \delta(L-x)$, in a similar way. This can be done, as $D_n(x)$ can also be written as

\begin{equation}
D_n(x)
=\sqrt \frac{2}{L} \sin(k_n  x)
= \sqrt \frac{2}{L} \sin[k_n  (x-L)] \cos (k_n L).
\label{Eq.TDE-6}
\end{equation}
Above, we have used $\sin(k_n L)=0$ and $\cos(k_n L)= \pm 1$. 
Using 
$\lim_{x\to L} \sin [k_n (x-L)]/(x-L) =k_n$, Then we have: 
$-\sqrt{2/L} \  k_n \cos(k_n L) \delta(L-x)= \ [D_n(x)/(L-x)] \delta(L-x)$. 
Finally,  Eq.~(\ref{Eq.TDE-5}) can be rewritten, together with the formula for $\Psi_n(x)$, as follows
\begin{eqnarray}
&& \Psi_n(x)= D_n(x) \theta(x) \theta(x-L),    
\label{Eq.TDE-7}\\
&& V(x)\Psi_n(x) 
=  \frac{\hbar^2}{2m}  \left[\frac{\delta(x)}{x}+ \frac{\delta(L-x)}{(L-x)\ }\right] D_n(x).
\label{Eq.TDE-8}
\end{eqnarray}
Dividing both sides of Eq.~(\ref{Eq.TDE-8}) by $D_n(x)$, we have
\begin{equation}
 V(x) \theta(x) \theta(L-x)
=  \frac{\hbar^2}{2m}  \left[\frac{\delta(x)}{x}+ \frac{\delta(L-x)}{(L-x)\ }\right].
\label{Eq.TDE-9}
\end{equation}
%
%
%We almost obtain the functional form of $V(x)$ of the infinite square well.
%Eq.~(\ref{Eq.TDE-9}) expresses the functional form of the infinite square well. 
Eq.~(\ref{Eq.TDE-9}) defines the infinite square well with a precise functional form. 
The factor $\theta(x) \theta(L-x)$ seems unavoidable.  
This factor is in fact needed, due to the divergence nature of the $V(x)$ outside the well. 
We may wonder how to describe the divergence in Eq.~(\ref{Eq.INT-1}).
%We may say this defines the infinite square well.
%
%
Eq.~(\ref{Eq.TDE-9}) may then be viewed as a specification formula for $V(x)$. Because  $V(x)$ is divergent outside the well, Eq.~(\ref{Eq.TDE-9}) shows that this divergence is eliminated when multiplied by the function $\theta(x) \theta(L-x)$.
Thus, the $\theta(x) \theta(L-x)$ term is needed to accompany with the $V(x)$ in order to have a regular result.
In other words, the $\theta(x) \theta(L-x)$ factor shows how to handle the infinity in the $V(x)$.
This is then interesting to note that to define the $V(x)$ of an infinite square well, the $V(x)$ can not be defined alone, it must be accompanied with the factor $\theta(x) \theta(L-x)$; a step better than using the notation $\infty$.

We may naively divide both sides of Eq.~(\ref{Eq.TDE-9})
by $\theta(x) \theta(L-x)$ to obtain a functional form of $V(x)$, but this is not formal, and in fact is not necessary.
The formula in Eq.~(\ref{Eq.TDE-9}) is enough for discussing the infinite square well model.
%
%
%A system defined in Eq.~(\ref{Eq.TDE-9}) requires that the wave function $\Psi(x)$ should be of the form as  $\Psi(x)= F(x) \theta(x) \theta(L-x)$; Then $V(x) \Psi(x)$ is well defined.
%
%otherwise $V(x) \Psi(x)$ is divergent if $\Psi(x)$ is not zero outside the well.
%
%
Multiplying both sides of Eq.~(\ref{Eq.TDE-9}) by an  arbitrary function $F(x)$, we obtain a regular form of $V(x) \Psi(x)$, with $\Psi(x)= F(x) \theta(x) \theta(L-x)$. 
Thus a particle confined inside an infinite square well with a wave function $\Psi(x)$, then the wave function and the potential energy should be described as follows 
\begin{eqnarray}
\label{Eq.TDE-10}
&& \Psi(x)= F(x) \theta(x) \theta(L-x),
\\
\label{Eq.TDE-11}
&& V(x)\Psi(x) =  \frac{\hbar^2}{2m}  \left[\frac{\delta(x)}{x}+ \frac{\delta(L-x)}{(L-x)\ }\right] F(x).
\end{eqnarray}
Eqs.~(\ref{Eq.TDE-10})-(\ref{Eq.TDE-11}) are a consequence of Eq.~(\ref{Eq.TDE-9}). Eq.~(\ref{Eq.TDE-10}) describes the wave function in an infinite square well, and Eq.~(\ref{Eq.TDE-11}) describes the functional form of the potential energy term in the Schr\"odinger equation. We see that $V(x) \Psi(x)$ is well defined. That $V(x) \Psi(x)$ is well defined is important, because it is $V(x) \Psi(x)$, not $V(x)$, that appears in the Schr\"odinger equation. A well defined $V(x) \Psi(x)$ then enables us to go on to solve energy eigenstates and eigenvalues of the Schr\"odinger equation, and  to calculate expectation values, etc.

We conclude that Eq.~(\ref{Eq.TDE-9}) defines the infinite square well. Or, equivalently, Eqs.~(\ref{Eq.TDE-10})-(\ref{Eq.TDE-11}) define the infinite square well. 
We call Eq.~(\ref{Eq.TDE-9}) or Eqs.~(\ref{Eq.TDE-10})-(\ref{Eq.TDE-11}) the reformed model of the infinite square well. We will see the advantage of this form in the following two sections. 

%
%

%%%%%%%%

\section{\label{sec:EIN} The energy eigenstates and eigenvalues of the reformed model }

We here solve the energy eigenstates and eigenvalues of the system defined in Eqs.~(\ref{Eq.TDE-10})-(\ref{Eq.TDE-11}). 
%Since Eq.~(\ref{Eq.TDE-9}), or Eqs.~(\ref{Eq.TDE-10})-(\ref{Eq.TDE-11}) are derived from $\Psi_n(x)$ and $E_n$, we expected the results should be the same as the original model.
The results can be obtained by the following two methods.

\textbf {Method (1)}\\

{case (I)} , $E>0$

We write the energy as $E= \hbar ^2 k^2 /(2m) $, where $k>0$. 
Substituting the wave function $\Psi(x)$ in  Eq.~(\ref{Eq.TDE-10}), and the $\Psi''(x)$ formulated in Eq.~(\ref{Eq.TDE-2}) into the Schr\"odinger equation in Eq.~(\ref{Eq.INT-2}), we then obtain an equation of the following form:

\begin{equation}
\label{Eq.EIN-1}
C_0(x)\ \theta(x) \theta(L-x) + C_1(x)\ \delta(x) + C_2(x)\ \delta(x-L)=0.
\end{equation}\\
where
\begin{eqnarray}
\label{Eq.EIN-2}
C_0(x) &=& \ F''(x) +  k^2 F(x).
\\
\label{Eq.EIN-3}
C_1(x) &=& \frac{F(x)}{x} -F'(x).
\\
\label{Eq.EIN-4}
C_2(x) &=& \frac{F(x)}{L-x} + F'(x).
\end{eqnarray}
Eq.~(\ref{Eq.EIN-1}) requires that $C_0(x)=0$, $ \lim_{x \to 0} C_1(x)=0 $ and $ \lim_{x \to L} C_2(x)=0 $. 
%%%%
%%%
From $C_0 (x)=0$, we obtain 

\begin{equation}
\label{Eq.EIN-5}
F(x) = A \sin(k x)+ B \cos(k x).
\end{equation}
Substituting this $F(x)$ into Eq.~(\ref{Eq.EIN-3}) yields
\begin{equation}
\label{Eq.EIN-6}
\lim_ { x\to 0} C_1(x)\  = \lim_ { x\to 0} \frac{B}{x}=0.
\end{equation}\\
Thus, it needs $B=0$. Therefore,
\begin{equation}
\label{Eq.EIN-7}
F(x)= A\ \sin(k x).
\end{equation}\\
Substituting Eq.~(\ref{Eq.EIN-7}) into Eq.~(\ref{Eq.EIN-4}) then yields
\begin{equation}
\label{Eq.EIN-8}
\lim_ { x\to L} C_2(x)\ = A \lim_ { x\to L}  \Big[ k \cos(k L)+ \frac{\sin(k x)}{L-x} \Big]=0.
\end{equation}\\
It then needs 
\begin{equation}
\label{Eq.EIN-9}
\sin(k L) =0.
\end{equation}
Eq.~(\ref{Eq.EIN-9}) implies $\cos(k L)= \pm 1$.
We can then use the identity, $  \sin(k x) = \sin(k (x-L))\cos(k L)$. This shows that the term, $k \cos(k L)+ \sin(k x)/(L-x)$ in the bracket of Eq.~(\ref{Eq.EIN-8}) is indeed approaching zero when $x$ approaches L.  

From Eqs.~(\ref{Eq.EIN-7})and (\ref{Eq.EIN-9}), we see that we have reproduced the solutions in Eq.~(\ref{Eq.INT-3}-\ref{Eq.INT-4}). We obtain these solutions by only using the precise form of  $V(x) \Psi(x)$; we need not put in by hand boundary conditions. That is, the precise form of  $V(x) \Psi(x)$ and the Schr\"odinger equation determine all. 

{case (II)} , $E=0$ 

In this case, $k=0$. From Eq.~(\ref{Eq.EIN-2}), we have $F(x)= A x +B$  \cite {b.r12}. 
%For the original model, we then have $ A=0$ and $B=0$ by the continuity of the wave function at the two sides. 
%However, we can also derive these results not by boundary condition. 
From  Eqs.~(\ref{Eq.EIN-3}), we have
$ \lim_{x \to 0} C_1(x)= \lim_{x \to 0}(B/x)=0 $, therefore $B=0$. And then from Eqs.~(\ref{Eq.EIN-4}), we obtain $ \lim_{x \to L} C_2(x)= \lim_{x \to L} A L/(L-x)=0 $. This requires that $A=0$. 
Thus, for $E=0$, we only have the trivial solution, $\Psi(x)=0$.

{case (III)} , $E<0$

In this case, $E=- \hbar ^2 k^2 /(2m) $. We then have $F(x)= A e^{k x}  +B e^{-k x}$. 
From  Eq.~(\ref{Eq.EIN-3}), we have
$ \lim_{x \to 0} C_1(x)= \lim_{x \to 0}(A+B)/x=0 $; therefore $B=-A$. And then from Eq.~(\ref{Eq.EIN-4}), we obtain $ \lim_{x \to L} C_2(x)= \lim_{x \to L} A [ \sinh(kL) /(L-x) + k \cosh(k L))]=0 $. This requires that $A=0$. 
Thus, for $E<0$, we again only have the trivial solution, $\Psi(x)=0$. \\

\textbf{Method-2}
\\
We may also solve this system by the usual way, that is, we separately solve $\Psi(x)$ in different regions and then connect these solutions at the boundaries. We consider here only the case for positive energy. 
From Eq.~(\ref{Eq.TDE-11}), $V(x)\Psi(x)=0$,
inside the well. We then easily obtain 
\begin{equation}
\label{Eq.EIN-10}
\Psi(x)= A \sin(k x)+B \cos(k x), \  0<x<L.
\end{equation}
We need to connect this solution with $\Psi(x)$ outside, which is $\Psi(x)=0$. The boundary condition can only be derived from the Schr\"odinger equation, which tells us the information about $\Delta \Psi'(x)$, where we let $\Delta \Psi'(x)$ represent the amount of change of $\Psi' (x)$ in the infinitesimal interval [$x-\epsilon,x+\epsilon$]. That is: $\Delta \Psi'(x)= \lim_{\epsilon \to 0} (\Psi'(x+\epsilon)-\Psi'(x-\epsilon))$. Thus
\begin{equation}
\label{Eq.EIN-11}
\Delta \Psi'(x)= \lim_{\epsilon \to 0}  \int_{x-\epsilon}^{x+\epsilon} \Psi'' (y)\  dy.  
\end{equation}
From Eq.~(\ref{Eq.INT-2}), $\Psi'' (y)$ can be replaced by $V(y)\Psi(y)$ and $E \Psi(y)$. The  $E \Psi(y)$ term makes no contribution to the integration. We then have 
\begin{equation}
\label{Eq.EIN-12}
\Delta \Psi'(x)=\lim_ {\epsilon\to 0}
\frac{2m}{\hbar^2}
 \int_{x-\epsilon}^{x+\epsilon} V(y) \Psi (y)\  dy. 
\end{equation}
Eq.~(\ref{Eq.EIN-12}) shows that the continuity of $\Psi'(x)$ is determined  by $V(x) \Psi(x)$.
For $x=0$, substituting Eqs.~(\ref{Eq.TDE-11}), (\ref{Eq.EIN-10}) into Eq.~(\ref{Eq.EIN-12}),
we then have
\begin{equation}
\label{Eq.EIN-13}
\Delta \Psi'(0)=\lim_ {x\to 0}(A k +\frac{B}{x}).
\end{equation}
Thus, it needs $B=0$. Then $\Psi(x)= A \sin(k x)$. Eq.~(\ref{Eq.EIN-13}) shows that  $ \Delta \Psi'(0)= A k$, therefore  $\Psi'(x)$ is not continuous at $x=0$.
For $x=L$, we have  
\begin{equation}
\label{Eq.EIN-14}
\Delta \Psi'(L)=\lim_{x\to L} \frac{A \sin(k x)}{L-x}.
\end{equation} 
It then needs  
\begin{equation}
\label{Eq.EIN-15}
\sin(k L)=0.
\end{equation} 
%
%\
And then $\Delta \Psi'(L)= - A k\ \cos(k L)$.

In conclusion, we obtain  $\Psi(x)= A \sin(k x)$, and the value of $k$ is determined by $\sin(k L)=0$.
We then have reproduce the solutions in 
Eqs.~(\ref{Eq.INT-3})-(\ref{Eq.INT-4}). It is interesting to note that, for the reformed  model, the boundary conditions derived are concerning about the values of $\Delta \Psi'(x)$, not $\Psi(x)$. 
And the values of $\Delta \Psi'(0)$ and $\Delta \Psi'(L)$ can be determined from Schr\"odinger equation.
In what follows, we discuss  ambiguities in the original infinite square well model, and we show that they can be resolved by the formula in Eq.~(\ref{Eq.INT-10}).
\section{\label{sec:RES} Resolving the ambiguities}
\subsection {\label{RES-1} The value of $V(x) \Psi_n (x)$ outside the well}

From Eq.~(\ref{Eq.INT-10}),  we have
$V(x) \Psi_n(x)=0$ outside the well. We thus resolve the ambiguity of $V(x) \Psi_n(x)= \infty \times 0=?$ in the original model. Actually, we see that 
$V(x) \Psi_n(x)=0$ everywhere, except at the two sides. 
The two delta functions in $V(x) \Psi_n(x)$ are important, discussed below. 

\subsection {\label{RES-2} The value of $V(x) \Psi_n (x)$ at the edges of the well}
From Eq.~(\ref{Eq.INT-10}), due to the delta function, we also have $V(x) \Psi_n(x)= \infty$ at the two edges of the well. The need of delta function at sides had been pointed out in Ref. \cite {b.r9, b.r10}. We thus also resolve the ambiguity of $V(x) \Psi_n(x)= \infty \times 0=?$ in the original model.
We can calculate the values of $\Delta \Psi_n'(0)$ and $\Delta \Psi_n'(L)$ from Eq.~(\ref{Eq.INT-10}).
By the solutions in Eq.~(\ref{Eq.INT-3}), we know that $\Psi'(x)$ is discontinuous at $x=0$ and $x=L$, and we obtain 

\begin{eqnarray}
&& \Delta \Psi_n'(0)=\sqrt{\frac{2}{L}}\ k_n ,
\label{Eq.RES-1a}\\
&& \Delta \Psi_n'(L)=-\sqrt{\frac{2}{L}}\ k_n \cos(k_n L). 
\label{Eq.RES-1b}
\end{eqnarray}
These results can not be checked from Eq.~(\ref{Eq.EIN-12}) for the original model, due to the vague notation $\infty$.
%
%This is because we would encounter the same  problem as  $V(y) \Psi(y) = \infty\times 0=? $ in the integration of $y$ over the  small interval $[x-\epsilon, x+\epsilon]$ for $x=0$ and $x=L$. 
%
%
%
For the reformed model, from Eqs.~(\ref{Eq.INT-10}), (\ref{Eq.EIN-12}), we  have:
\begin{eqnarray}
\label{Eq.RES-2}
\Delta \Psi_n' (0)&=& \frac {2m}{\hbar^2}\ \int_{-\epsilon}^{\epsilon} \sqrt{\frac{2}{L}}\ \frac{\hbar^2}{2m}\ k_n \  \delta(y) dy= \sqrt{\frac{2}{L}}\ k_n,
\\
\nonumber\\
\label{Eq.RES-3}
\Delta \Psi_n' (L)&=& \frac {2m}{\hbar^2}\ \int_{L-\epsilon}^{L+\epsilon} \sqrt{\frac{2}{L}}\ \frac{\hbar^2}{2m}\ k_n \ [ - \cos(k_n L)  \delta(L-y)] dy
\nonumber\\
&=& - \sqrt{\frac{2}{L}}\ k_n \cos(k_n L).
\end{eqnarray}
These results in Eqs.~(\ref{Eq.RES-2}-\ref{Eq.RES-3}) are consistent with those in Eqs.~(\ref{Eq.RES-1a})-(\ref{Eq.RES-1b}).
From Eqs.~(\ref{Eq.RES-2}-\ref{Eq.RES-3}) , we note that if there were no delta functions, but only ordinary functions in $V(x) \Psi_n(x)$, then $\Delta \Psi_n' (0)=0$
 and $\Delta \Psi_n' (L)=0$. That means $\Psi'(x)$ is continuous at the two sides of the well, and we will  only obtain a trivial solution, $\Psi(x)=0$.  The two delta functions in the $V(x) \Psi_n(x)$ in  Eq.~(\ref{Eq.INT-10}) are therefore important, as they change a trivial solution to a non-trivial solution.

\subsection {\label{RES-3} The manifestation of Ehrenfest's theorem}
The third puzzle concerns about the manifestation of Ehrenfest's theorem 
for time-evolved wave packets in the infinite square well. The time evolution of a general wave packet $\Psi(x,t)$ is as follows
\begin{eqnarray}
\label{Eq.RES-4}
\Psi(x,t)=\sum_{n=1}^{\infty} a_n \Psi_n(x) e^{- i  \omega_n t}. 
\end{eqnarray}
where $\omega_n=E_n/\hbar$. 
To verify Ehrenfest's theorem, we need to verify the following formula:
\begin{equation}
\label{Eq.RES-5}
\frac{d}{dt} \langle\Psi(t) \mid \hat{p} \mid \Psi(t)\rangle  =  -\langle\Psi(t) \mid \frac{dV(\hat{x})}{d\hat{x}} \mid \Psi(t)\rangle.
\end{equation}
The calculation of the right side of  Eq.~(\ref{Eq.RES-5}) is related to the calculation of $\Psi_n(x)  (dV(x)/dx) \Psi_j(x)$. We note that
\begin{eqnarray}
&& \Psi_n(x)  \frac{dV(x)}{dx} \Psi_j(x)
\nonumber\\
&& = \frac{d}{dx}  [\Psi_n(x) V(x) \Psi_j(x)]-
\frac{d \Psi_n(x)}{dx}   [V(x) \Psi_j(x)]-
[\Psi_n(x) V(x)] \frac{d \Psi_j(x)}{dx}.
\label{Eq.RES-6}
\end{eqnarray}
For the right side of Eq.~(\ref{Eq.RES-6}), the first term makes no contribution in the calculation of integration.
Using Eq.~(\ref{Eq.INT-10}), the other two terms can be calculated, and we obtain
\begin{equation}
\frac{d \Psi_n(x)}{dx} V(x)\Psi_j(x)
=
\frac{\hbar^2}{m L}\ k_n k_j\ \theta(x) \theta(L-x)\ [\delta(x)-(-1)^{n+j} \delta(L-x)]. 
\label{Eq.RES-7}
\end{equation}
And then we have
\begin{eqnarray}
&&\int_{-\infty}^{\infty} \frac{d \Psi_n(x)}{dx} V(x)\Psi_j(x) dx
\nonumber\\
&&=\frac{\hbar^2}{m L}\ k_n k_j \int_{0}^{L} [\delta(x)-(-1)^{n+j} \delta(L-x)] dx
\nonumber\\
&& =\frac{\hbar^2}{2 m L}\ k_n k_j\ \beta_{n j}.
\label{Eq.RES-8}
\end{eqnarray}

where $\beta_{n j}=1-(-1)^{n+j} $. Above, we have used the following results

\begin{eqnarray}
&&\int_{0}^{L} \delta(x)\ dx = \frac{1}{2},
\label{Eq.RES-9}
\\
&&\int_{0}^{L} \delta(L-x)\ dx = \frac{1}{2}.
\label{Eq.RES-10}
\end{eqnarray}
These results are due to the even function property of the delta function, so that the integration of $x$,  over the range, $[0, L]$, is equal to one half of that over the range, $[-L, L]$.
We whence obtain the final result
\begin{eqnarray}
\label{Eq.RES-11}
&&\langle\Psi(t) \mid \frac{dV(\hat{x})}{d\hat{x}} \mid \Psi(t)\rangle 
\nonumber\\
 &=&\int_{-\infty}^{\infty} \Psi^*(x,t) \frac{dV(x)}{dx} \Psi(x,t) dx 
 \nonumber\\
 &=&-\frac{\hbar^2}{m L}\sum_{n=1}^{\infty} \sum_{j=1}^{\infty}
 a_n^* a_j k_n k_j \beta_{n j} e^{i (\omega_n-\omega_j)t }.
\end{eqnarray}
We also have
\begin{eqnarray}
\label{Eq.RES-12}
&&\langle\Psi(t) \mid \hat{p} \mid \Psi(t)\rangle  \nonumber  \\
 &=&(- i \hbar)\ \frac{2}{ L}\ \sum_{n=1}^{\infty} \sum_{j=1, j\neq n}^{\infty}
 a_n^* a_j \frac{k_n k_j}{k_n^2-k_j^2} \beta_{n j} e^{i (\omega_n-\omega_j)t }.
\nonumber\\
\end{eqnarray}
And then, we have
\begin{eqnarray}
\label{Eq.RES-13}
  &&\frac{d}{dt}\langle\Psi(t)\mid \hat{p} \mid \Psi(t)\rangle 
  \nonumber\\
 &=& \frac{\hbar ^2}{ m L}\ \sum_{n=1}^{\infty} \sum_{j=1}^{\infty}
 a_n^* a_j k_n k_j \beta_{n j} e^{i (\omega_n-\omega_j)t }.
\end{eqnarray}
Comparing Eqs.~(\ref{Eq.RES-11}) and (\ref{Eq.RES-13}), we see that Ehrenfest's theorem is confirmed. 
The calculations are done directly using the function $V(x) \Psi_n(x)$ in Eq.~(\ref{Eq.INT-10}). 
We need not first do those calculations in the finite square well and then take the limit, $V_0 \to \infty$  \cite{b.r12}. It had been argued that the infinite potential can not be simply described as the limit of a finite one \cite{b.r13}. 
The potential energy $V(x)$ in a finite well is well defined, see Eq.~(\ref{Eq.App-1}). Simply taking the limit $V_0 \to \infty$ results the potential energy $V(x)$ defined in Eq.~(\ref{Eq.INT-1}), which does not have a well defined functional form. It turns out that the limit should not be taken in the potential energy $V(x)$. Instead, our results show that the limit should be taken in the potential energy term $V(x) \Psi(x)$.

\section{\label{sec:C} Conclusion}

Infinite square well model is a model using infinite large potential barrier to confine particles inside a well. 
%
%
%In this paper, we show that, instead of using the notation $\infty$ and  the imposed boundary conditions, we can derive a precise functional form for the infinite square well. 
In this paper, we show that we need to avoid using the notation $\infty$, because it requires imposing extra boundary  condition and it also brings in ambiguities. 
Somehow this means that using the notation $\infty$ does not define the system precisely; there is some degree of freedom remained, and it
needs extra boundary condition to fix the solution.

To investigate a more precise form of the potential energy, we start from the well-known solutions of the infinite square well, and go back to determine  the corresponding potential energy.
%
%we show that we have derived a precise functional form for the infinite square well, which is described in Eq.~(\ref{Eq.TDE-9}). 
%
%
The derived potential energy $V(x)$ is described in Eq.~(\ref{Eq.TDE-9}). We can have an understanding of the divergence of the $V(x)$ by noting that the $V(x)$ is needed to be  accompanied with the function $ \theta(x) \theta(L-x)$.
%
%
%Infinite square well model is a model using infinite large potential barrier to confine particles inside a well, instead of using the notation $\infty$ and  the imposed boundary conditions, we can derive a precise functional form for the infinite square well.
%
%
In another point of view, what we have done is to showing the specification of the infinity in Eq.~(\ref{Eq.INT-1}) such that it results the known solutions of $\Psi_n(x)$ and $E_n$.
That is, the derived functional form of the $V(x)$ gives a specification or a description about how to handle the infinity in the original model.

Such an infinite large barrier, however,  is not enough to confine particles inside a well.  To confine particles inside a well, we also need delta functions, $\delta(x)$ and  $\delta(L-x)$ in the form of $V(x)$. These two delta functions are needed for constructing non-trivial solutions. 
%%
%%

%
%
%Infinite square well is subtle, the potential energy defined in Eq.~(\ref{Eq.TDE-14}) may be as vague as the notation $\infty$ in the original model defined in Eq.~(\ref{Eq.INT-1}). However, for the reformed model, the potential energy term, $V(x) \Psi(x)$, has a precise functional form. The original model does not show this property.

%We show that a formal functional form  for an infinite square well is defined via $V(x) \Psi(x)$. This form is described in Eqs.~(\ref{Eq.TDE-16})-(\ref{Eq.TDE-17}). Schr\"odinger equation is then well defined, though the form of $V(x)$ may not be so a formal one. This shows that constructing a confining potential is subtle. 

%
%
%
We show that the energy eigenfunctions and eigenvalues of this reformed model are the same as the original ones. The difference between the original and the reformed model in obtaining solutions is that, for the reformed model, we need not impose boundary conditions at the two sides, we need only use the precise functional form of $V(x) \Psi(x)$. 
Readers, who are interested in other type of solutions of the infinite square well, in which the continuity of $\Psi(x)$ at boundaries is not required, may refer to Ref. \cite{yshuang}. 

Besides without the need of imposing boundary conditions, we show that the reformed model enables us to calculate quantities directly and precisely, such as the values of $\Delta \Psi'(0)$, $\Delta \Psi'(L)$, the expectation values $ \langle\Psi(t) \mid d\hat{V} (x)/dx \mid \Psi(t)\rangle  $, and Ehrenfest's theorem can be confirmed directly. 
%
%%\%%
%%
Finally, for discussing the time evolution of the wave function $\Psi(x,t)$ in the reformed model, we need only replace the $F(x)$ in Eqs.~(\ref{Eq.TDE-10})-(\ref{Eq.TDE-11}) by $F(x,t)$.
\\
\\
\\

%%%%%%%%
%%%%%%%%

 \begin{center}
 {\bf APPENDIX I}
 \end{center}
 
In this appendix, we drive the functional form of the potential energy $V(x)$ for a finite well from the known solutions of $\Psi(x)$.
The potential energy $V(x)$ of the finite square well is described by
\begin{equation}
\label{Eq.App-1}
V(x)=\left\{
\begin{array}{ll}
0,&        0<x<L, \\
V_0,&      \text{otherwise}, 
\end{array}
\right.
\end{equation}
where $V_0 >0$. We set the energy $E =\hbar^2  k^2/ 2 m $, and let  $V_0 -E =\hbar^2  q^2/ 2 m $.  The wave number  $k > 0$ is determined by:
\begin{equation}
\label{Eq.App-2}
\ \tan(k L)=\frac{2 k q}{k^2-q^2}. 
\end{equation}
or,
\begin{equation}
\label{Eq.App-3}
\left \{
\begin{array}{ll}
\sin(k L)= a\ \frac{2 k q}{k^2+q^2},&      \\
\cos(k L)= a\ \frac{ k^2-q^2}{k^2+q^2},&     
\end{array}
\right.
\end{equation}
where $a=\pm 1$; which value of $a$ should be chosen depends on the value of $k L$ in Eq.~(\ref{Eq.App-3}). The energy eigenfunction is known to be 

\begin{equation}
\label{Eq.App-4}
\Psi(x)=\left\{
\begin{array}{ll}
A\   e^{q x},&        x<0,\\
A\ D(x), &    0<x<L, \\
a\ A\ e^{-q x},&      x>L. 
\end{array}
\right.
\end{equation}
where $A$ is the normalization constant, and $D(x) =  \frac{q}{k} sin(k x) + \cos(k x)$. We have

\begin{equation}
\label{Eq.App-5}
A =  \sqrt { \frac{2}{L} }  \frac{k}{q}  {\frac{1}{
\sqrt{(1+\frac{2}{q L}) (1+ \frac{k^2}{q^2})} }
} .     
\end{equation}
We now write the $\Psi(x)$ in Eq.~(\ref{Eq.App-4}) as:
\begin{equation}
\label{Eq.App-6}
\ \Psi(x)= A\ [ e^{ q x}\theta(-x)+ D(x)\ \theta(x)\ \theta(L-x)+ a\ e^{-q x} \theta(x-L)].\\    
\end{equation}
To determine $V(x)$ from the Schr\"odinger equation in Eq.~(\ref{Eq.INT-2}), we rewrite the equation as 
\begin{equation}
\label{Eq.App-7}
V(x) \Psi(x)= \frac{\hbar^2}{2m}\Psi'' (x)+E \Psi(x). 
\end{equation}
Substituting the $\Psi(x)$ in Eq.~(\ref{Eq.App-6}) into the right side of Eq.~(\ref{Eq.App-7}), we can then read out the form of $V(x) \Psi(x)$. After some calculation, we obtain 
\begin{eqnarray}
\ V(x) \Psi(x) &=& V_0\ \theta(-x) A\ e^ {q x} + V_0\ \theta(x-L)\ a\ A\ e^{-q x}. 
\nonumber\\
\label{Eq.App-8}
&=& V_0\ \theta(-x)\Psi(x) +  V_0\ \theta(x-L)\Psi(x) 
\end{eqnarray}
This yields that
\begin{equation}
\label{Eq.App-9}
\ V(x) = V_0\ \theta(-x) + V_0\ \theta(x-L). \\  
\end{equation}
This is just the potential energy defined in  Eq.~(\ref{Eq.App-1}). Thus, we have  derived $V(x)$ from $\Psi(x)$.

\begin{acknowledgments}
The author is indebted to Prof. Young-Sea Huang for bringing in the attention and interests of this subject, and also for much help in discussions. The author would also like to thank Prof. Tsin-Fu Jiang and Prof. Wen-Chung Huang for much help. 
\end{acknowledgments}

\end{document}